# Solving the Tyranny of Pipetting

*Stephen Quake - Stanford University*

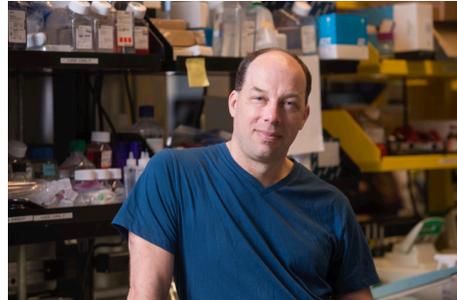

### Biography

*Stephen Quake studied physics (BS 1991) and mathematics (MS 1991) at Stanford University, where he was elected to Phi Beta Kappa and won the Firestone Prize for Undergraduate Research. His senior thesis research on the manipulation of single DNA molecules with optical tweezers won the Apker Award from the American Physical Society, an honor given to the best undergraduate physics research nationwide. Quake also won a British Marshall scholarship and a National Science Foundation graduate fellowship, which he used to earn a doctorate in theoretical physics from Oxford University (DPhil 1994). His thesis research was in statistical mechanics and the effects of knots on polymers. He then spent two years as a post-doc in Nobel laureate Steven Chu's group at Stanford University continuing his research in single molecule biophysics.*

*Quake joined the faculty of the California Institute of Technology in 1996, where he rose through the ranks and was ultimately appointed the Thomas and Doris Everhart Professor of Applied Physics and Physics. At Caltech, Quake received "Career" and "First" awards from the National Science Foundation and National Institutes of Health, and was named a Packard Fellow. These awards supported a research program that began with single molecule biophysics and soon expanded to include the invention of microfluidic large scale integration and its application to biological automation, the first microfluidic droplet devices, and the first single molecule DNA sequencing. He moved back to Stanford University in 2005 to help launch a new department in Bioengineering, where he is now the Lee Otterson Professor of Bioengineering and Applied Physics.*

*At Stanford Quake won one of the inaugural NIH Director's Pioneer Awards and was an investigator of the Howard Hughes Medical Institute for a decade. His research in microfluidics continued during this period, and among many other discoveries he developed approaches for single cell genome and transcriptome analysis, which he and collaborators used to investigate a variety of questions in cell biology. He also began using genomic approaches to create new diagnostic tools, including the first non-invasive prenatal test for Down syndrome and other aneuploidies. His test is rapidly replacing risky invasive approaches such as amniocentesis, and millions of women each year now benefit from this approach. His innovations have helped to radically accelerate the pace of biology and have made medicine safer by replacing invasive biopsies with simple blood tests. In 2016 Quake became the founding co-president of the Chan Zuckerberg Biohub, a non-profit medical research organization which he leads with Joe DeRisi, while also maintaining an active research lab at Stanford.*

*Quake is an elected fellow of the National Academy of Sciences, the National Academy of Engineering, the National Academy of Medicine, the National Academy of Inventors, the American Physical Society and the American Institute for Medical and Biological Engineering. He is a recipient of numerous international awards, including the Raymond and Beverly Sackler Prize for Convergence Research, The Max Delbruck Prize in Biological Physics, the Gabbay Prize for Biotechnology and Medicine, the Human Frontiers of Science Program Nakasone Prize, the MIT-Lemelson Prize for Inventors, the Raymond and Beverly Sackler International Prize in Biophysics, the American Society of Microbiology's Promega Biotechnology Award, and the Royal Society of Chemistry Publishing's Pioneer of Miniaturization Award.*

*Quake has published more than 225 research papers in numerous leading scientific journals, including Science, Nature, Cell and P.N.A.S. His publications have been cited more than 25,000 times and his h-index is 72. His paper describing the invention of the micromechanical valve that forms the basis of microfluidic large scale integration has been cited more than 2,000 times, and his papers describing both the first microfluidic large scale integration as well as the first microfluidic droplet devices and their associated physics have each been cited more than 1,000 times. The practical applications of his research have led to more than 130 issued U.S. patents, as well as numerous international patents. Quake is an active entrepreneur who has founded or co-founded several companies, including Fluidigm, Helicos, Verinata Health, Moleculo, Immumetrix, Quanticel Pharmaceuticals, Cellular Research, Karius, Agenovir and Molecular Stethoscope.*

## Introduction

*"There is no such thing as a special category of science called applied science; there is science and its applications, which are related to one another as the fruit is related to the tree that has borne it."* - Louis Pasteur

Fluid mechanics is the study of flow, and it is a broad topic that includes all phases of matter, from liquid to gas to plasma – even comprising solids, in the cases of fluidized beds and granular materials such as sand. The basic equations as defined by Navier and Stokes amount to a statement of conservation of momentum and mass, and yield a fantastically rich collection of solutions and phenomena. Much attention in the 20$^{th}$ century focused on understanding turbulent behavior at large length and energy scales, but it turns out that there are equally fascinating effects to understand at small length and energy scales. This field, which is called "microfluidics", has its origins in the 1970's and continues to the present day to be a rich source of ideas and discoveries, both technologically and scientifically, and has led to a large number of important industrial applications. Here I will review the early history of the field and discuss more recent developments, with a focus on applications in biology and biochemistry.

## The Technology of Microfluidics

A key moment in the technological origin of microfluidics is generally considered to be the Stanford gas chromatograph developed by micromachining pioneer James Angell and his student Stephen Terry in 1979.[1,2] This was a fully integrated device that included a column, detector, and micromechanical valve to regulate gas flow. Terry et al were able to demonstrate chromatographic performance and even calculated the resolution in theoretical plates. They contemplated applications ranging from use in space missions to biological implants. This pioneering work captured the imagination of researchers around the world and inspired much future activity in the field, but did not lead directly to commercial development or use. In contrast, the development of microfluidic ink jet print heads during the same period[3,4] led to broad commercial success and widespread use in the personal computer industry. Ink jet printers were the first widely used microfluidic products and have in their own right inspired numerous spin-off applications, including in biochemical synthesis.

Both the gas chromatograph and the ink jet printer were fabricated using silicon micromachining techniques, and it is important to note the relationship between new fabrication approaches and the development of microfluidic technologies. Initially, such devices were made from silicon using techniques originally developed for the fabrication of electronic integrated circuits. James Angell was the first to coin the term "micromachining" and his work led to the genesis of a field (also called microfabricated electromechanical machines, or MEMS) which created accelerometers and many other mechanical devices. It is one of the pleasant surprises of science that silicon, which was intensively studied and developed for its electronic properties, also turns out to be an excellent mechanical material.[5]

The field of microfluidics began to develop rapidly after the popularization of Angell's gas chromatograph. A number of analytical chemists, both in industry and in academia, enthusiastically pursued the vision of an integrated "lab on a chip". At Ciba Geigy, the chemist Andreas Manz hosted Jed Harrison on a sabbatical visit and together they developed a microfabricated capillary electrophoresis device.[6] Michael Ramsey at Oak Ridge National labs independently developed similar microfabricated separations devices.[7] The work of these groups was commercially developed by a startup company called Caliper Technologies, which created products for electrophoresis of DNA, RNA and proteins. Meanwhile, a team at Pharmacia in Uppsala, Sweden spun out a "biosensor" group in the form of Biacore which developed a surface plasmon sensor with microfluidic sample manipulation.[8] This work led to the eponymous Biacore machine for measuring molecular affinities, which has become widely used in the pharmaceutical industry. Several other companies joined the effort and a thriving academic community has pursued a number of applications of microfluidics to analytical chemistry using related approaches.[9,10]

Despite the commercial success of Biacore and Caliper, those microfluidic technologies could only perform specific tasks, were limited in flexibility and did not achieve the elusive dream of a "lab on a chip". A key reason for this was that they had not solved the general problem of how to achieve arbitrary manipulation of fluids on a chip. Electrokinetic techniques such as used by Caliper are useful for simple manipulations, but do not generalize to complex networks because it is not practical to perfectly balance a large network of effectively linear resistors, and all of the fluidic junctions are susceptible to diffusional leakage and undesired flows due to temperature changes and imperfections in fabrication. In principle, these difficulties can all be overcome by the use of mechanical valves, and indeed there were several examples of microfabricated mechanical valves in the literature dating from Angell's chromatograph - but none of those valves could be fabricated in a fully integrated and scalable fashion.[11–14]

The difficulty and technical gap can be best understood through an analogy with electronics. The development of discrete components such as vacuum tubes and transistors enabled the development of powerful and useful electronic devices, including the first electronic computers. Electrical engineers realized that they could design circuits of arbitrary complexity – but that there was a limit to the complexity of circuits which could be built in practice, due to errors in assembly, cold solder joints, and component failures. They called this limit the "tyranny of numbers", and it represented a similar limitation for circuit design as was faced by the microfluidics community.[15] In electronics, this problem was solved by Jack Kilby and Robert Noyce with their invention of the integration circuit and large scale integration, which enabled batch fabrication of arbitrarily large numbers of components with a small number of lithographic, deposition and processing steps. It is important to note that Noyce and Kilby did not invent the first transistor; rather, they invented a way to fabricate large numbers of transistors and other electrical components which in turn revolutionized the complexity of electronic circuits which could be practically fabricated.[16]

The development of an equivalent large scale integration in microfluidics awaited the birth of a new revolution in microfabrication – the development of "soft lithography" by George Whitesides in the late 1990's.[17] Whitesides realized that the focus by MEMS and microfluidics researchers on hard materials such as silicon and glass was unnecessarily myopic. In fact, the world of chemistry offered a much broader palette of materials to fabricate with which could unleash fantastic creativity. The origin of Whitesides's insight dates back to the 1940's, when it was first realized that polymer materials could be used to mold replicas of very small sub-micron features.[18] This observation was made in the context of electron microscopy, and the technique became widely used as a tool for enhancing contrast of three dimensional features for imaging. Whitesides adopted this replica molding approach for the purposes of chemical printing, by fabricating featured "stamps",[19] and then realized that it could be used much more generally for fabricating a wide variety of structures that were simply not accessible through planar silicon micromachining. His group went on to fabricate a dizzying array of structures, including linked rings,[20] three dimensional conducting coils,[21] basket weave structures,[22] blazed gratings,[23] lasers,[24] and optical wave guides.[25] Soft lithography was eventually applied to make functional microfluidic structures, of which the earliest examples were for electrophoresis, molecular cytometry, and cell sorting.[26–29] Soft lithography became widely adopted in the academic community and has become the de facto fabrication standard for many applications of microfluidics, especially in biology.[30,31] It is difficult to overstate the impact this approach has had on the field of microfluidics due to its simplicity, flexibility, and rapid turnaround time – all of which empower researchers in ways that weren't possible previously.

Inspired by the power of soft lithography and the flexibility (both literally and metaphorically) of polydimethylsiloxane, the preferred material which Whitesides had also popularized, my group invented a monolithic integrated valve which could be batch fabricated.[32] We realized that every macroscopic valve needs a flexible rubber valve seat to prevent leaking (usually supplied by a rubber washer) and we decided to fabricate the entire valve out of the flexible material PDMS. By bonding together two layers of molded PDMS structures produced by soft lithography (so-called "Multilayer Soft Lithography"), we were able to fabricate pairs of perpendicular channels in separate layers, which were separated only by a thin PDMS membrane at the point at which they crossed each other. When we applied pneumatic pressure to one of the channels, the membrane deflected into the other channel, thus blocking fluid flow.  Making a leak-proof valve required us to develop approaches which enabled careful control of the channel geometry of the mold, since rectangular cross-section channels would not seal perfectly due to the singularity required in membrane deflection.  Thus we adopted a process to create rounded channels on the mold structure and hence conformal deflection of the membrane would create a leak-proof seal.  Another technical innovation that was required was to understand and compensate for differential shrinkage of PDMS during the curing process, since the two layers had slightly different chemical composition and hence difference curing properties.  Finally, we had to figure out how to bond two PDMS layers together, which we accomplished by using off-ratio mixing of the two components, creating an excess of monomer in one layer and an excess of catalyst in the other layer.

The beauty of this approach is that fabrication is simple – two lithography steps to create to separate molds, followed by two PDMS molding steps, followed by a bonding step.  Since the features were determined by lithography, we could make as many valves as we wanted in a single device without increasing the fabrication complexity – we only had to design a more complicated mask on the computer.  We demonstrated this rather dramatically by fabricating devices with hundreds of valves in a single chip and showed that we could create microfluidic memory devices, as well as devices used for screening cellular reactions.  We termed this approach "Microfluidic Large Scale Integration", in analogy with microelectronic LSI, and realized we had solved the outstanding problem that had prevented realization of the "Lab on a Chip" vision.  This breakthrough was published in 2002,[33] a mere two years after the first publication of our valve technology, and represented two orders of magnitude increase in complexity per chip.  These integrated valves have found many applications in biology, including protein crystallization for structural biology, measuring molecular affinities, digital PCR, and single cell genomics.  They have been commercially developed and are manufactured by Fluidigm, whose products are now used in thousands of labs around the world.  Fluidigm's most complex chips contain tens of thousands of valves each, and cumulatively Fluidigm has manufactured and shipped billions of valves to customers.  Collectively, this shows that PDMS chips are manufacturable and that chips with high density valves have been successfully transitioned from academia to routine commercial use, thus completing the vision of microfluidic automation of biology as an analogue of electronic automation of computation some two decades after it was first articulated.

It is natural to ask how to move beyond thousands of experiments on a chip to millions and larger.  Valve densities have improved according to a microfluidic "Moore's law",[34] but in order to keep improving experimental density and throughput one will probably need to develop other technologies.  A promising candidate for this is microfluidic emulsions on a chip, or "droplet microfluidics", first developed in my lab in 2001.[35]  We were inspired by a paper by Dan Tawfik and Andrew Griffiths, who showed that in certain special cases benchtop emulsions could be used to perform laboratory evolution of proteins.[36]  The emulsions were used to encapsulate separate reactions and measure the product from a given protein mutant.  Unfortunately, this approach placed severe restrictions on the type of assay one could use, and we realized that if we could create a microfluidic emulsion format, it would be possible to screen each droplet individually, thus vastly increasing the flexibility of the system and the types of proteins that could be evolved.  We designed chips that had junctions of two channels into a third output channel, and filled one of the inputs with water and the other with oil.  Much to our surprise, when we applied constant pressures on the inputs, an instability developed at the junction and small droplets of water were periodically budding off into the flowing stream of oil, thus naturally creating emulsions in large amounts without the need for external manipulation on a per droplet basis.

As we began to study the system, we realized that there was a rich variety of physical behaviors. We could make polydisperse emulsions, ribbons of droplets with long range coherent order, trains of droplets and so forth. We spent some effort trying to understand these and explain the basic physics of the system,[35] and our first publication attracted interest in the fluid physics community.[37] We made a variety of devices both to explore the basic droplet formation phenomena and also to turn it into a screening system for molecular evolution, including showing that we could sort droplets with integrated valves on the chip, that we could trap bacteria in the droplets, and that we could assay enzyme kinetics in the droplets.[38] This idea of using droplets to screen experiments was enthusiastically accepted by the microfluidics community and a number of researchers jumped into the field, making rapid progress.[39] Rustem Ismagilov showed that our work in protein crystallization could be transported to the droplets, that one could assay various enzyme kinetics in a rigorous fashion, and that one could study a variety of complex chemical phenomena.[40] David Weitz developed a way to sort droplets at high speed using integrated electrodes and decided to complete the program of screening for molecular evolution of proteins, with dramatic success.[41] BioRad offers a digital PCR system based on droplet microfluidics. In addition to several companies now offering products based on this technology, there is a thriving academic community which continues to explore the basic science and technological applications of droplet microfluidics. This approach enables very large numbers of experiments on a chip, but without the sophisticated fluid manipulation flexibility achieved with the valves, and is often a valuable complementary approach.

## The Science of Microfluidics

There is a rich and diverse physics to explore in trying to understand the behavior of fluids in small volumes (or at small length scales). Often these phenomena are organized according to dimensionless numbers which estimate the relative magnitude of competing effects. The most famous of these numbers is the Reynolds number, which was introduced to understand the transition from laminar flow to turbulence, and which estimates the ratio of inertial forces to viscous (ie friction) forces. At small length scales, the Reynolds number is small, which means that flow is laminar and dominated by viscous friction. This is not a small effect: the practical consequences of this on one's ability to swim were famously discussed by Ed Purcell in his monograph "Life at Low Reynolds Number".[42] There are many other physical effects that make the microfluidic world quite different from our experience with macroscopic fluids, and there has been a tremendous amount of research and cleverness devoted to understanding these effects and exploiting them to create microfluidic devices without a macroscale analogue, as described here.[43]

Early critics of the field argued that microfluidics was "trivial" because low Reynolds number renders the differential equations describing fluid flow linear, resulting in "simple" Stokes flow. However, the situation is actually much more subtle than that, and considering microfluidic problems has actually been quite fertile in terms of understanding basic fluid physics. As a simple example, laminar flow presents a well controlled experimental environment to explore the behavior of complex fluids such as polymer solutions. The dimensionless number associated with a polymer solution is the Weissenberg number, which is the product of the intrinsic relaxation time of the polymer with the shear rate of the fluid, and it is an estimate of the energy being put into stretching the polymer. It is often challenging to understand complex fluids at the macroscale because the non-linearities of inertial flow get convolved with the non-linearities of the polymer solution. However, studying complex fluids in microscale devices leaves one with only the nonlinearity of interest to study, because one can simultaneously be at low Reynolds number and high Weissenberg number. We used this insight to create devices which displayed a variety of interesting behaviors in the realm of "fluidic logic", including a constant current source,[44] a fluidic rectifier,[45] and a fluidic flip-flop gate,[44] all of which were specially designed channel geometries filled with a dilute polymer solution. These devices provide a route to creating miniaturized fluidic logic circuits since macroscale fluidic logic depends on inertial effects which do not scale usefully.

For many years the low Reynolds number environment was considered a major challenge for microfluidics: how could one achieve efficient mixing of fluids when transport is dominated by diffusion, which is generally a slow process? Many groups ultimately found clever solutions to problems, and one of the most powerful and useful ideas came from Whitesides, who showed that it was possible to create a "chaotic mixer" which used patterned surfaces to create circulating flows which cause an advecting fluid to fold over itself in a baker's dough pattern.[46] This reduces the length scales over which diffusion needs to act, and in turns greatly accelerates mixing in a fluid as it flows through the channel. Many groups have made use of this component in their microfluidic devices. Whitesides also took advantage of the properties of low Reynolds number flow to design a microfluidic gradient generator, which allows one to create chemical gradients of arbitrary shape in a flowing fluid based on clever microfluidic channel design.[47] They showed that it was possible to use this device to create gradients of a variety of shapes, and that such gradients could be used to study chemotaxis of cells.[48]

Another example of how small volume fluid physics can be put to use is found in protein crystallization. The most widely used technique to determine the structure of proteins is x-ray crystallography, and this technique requires the protein to be grown in an atomically ordered crystal. This can be challenging both because it is impossible to predict the correct protein crystallization conditions, necessitating many experiments, and because it is in general expensive and challenging to produce large amounts of a given protein. Both of these challenges make protein crystallization a natural topic for microfluidic automation, and as my group began experimenting with microfluidic crystallization we noticed that we were having surprising success in our ability to grow crystals.[49] As we started to think about what might be happening in our chip relative to bench top experiments, we realized that our crystallization experiments were operating in an interesting fluidic regime – that of low Grashof number. The Grashof number is defined as the ratio of buoyant forces to viscous forces, and therefore density driven convection is negligible when the Grashof number us small. The particular geometry of the chambers in our experiment and the diffusion-driven mixing kinetics of protein and precipitant (in conjunction with the absence of convection) meant that our experiments were implementing a crystallization reaction known as free interface diffusion (FID). FID represents a heuristically optimal traversal of phase space, in that proteins see a rapidly increasing precipitant concentration, which serves to nucleate crystals, followed by a decreasing protein concentration, which slows the growth kinetics to allow atomically ordered crystals to form. This particular approach is in general challenging to implement, and was a motivation for NASA to perform microgravity crystallization experiments in space, but it becomes absolutely straightforward to do on earth in a microfluidic environment.

## Microfluidics in Biology and Biochemistry

Microfluidic technologies have found applications in many fields – including medical diagnostics, ink jet printers, analytical chemistry, batteries and energy storage, and water filtration and purification. In this section I will focus on some of the substantial and transformative applications in one of the most exciting and fast moving fields of science: biology. Microfluidic devices provide a powerful approach for automation and have enabled experiments that would not otherwise be possible in areas as diverse as structural biology, biochemistry, biophysics, genetics, and cell biology. This has been a focus of my own research and after having developed microfluidic large scale integration, we then set out to explore its applications to biology.

One of the first areas we explored was structural biology, for the reasons outlined above. In collaboration with James Berger we developed a chip to do highly parallel FID crystallization experiments – 144 experiments with only a few microliters of protein solution.[49] This chip was commercialized by Fluidigm and was used in structural biology labs around the world for nearly a decade. Many of the structures were determined by pharmaceutical companies and kept proprietary, but a number of structures solved by academic groups were published in the scientific literature.[50–61] Some of these structures were very important in human health, and include an Ebola virus glycoprotein bound to an antibody from a survivor,[57] hemagglutinin from from H5N1 influenza virus,[58] the LDL cholesterol receptor complexed with PCSK9,[59] and integrins bound to fibrinogen-mimic therapeutics.[60] Other chips we developed for the purpose of structural biology include a "formulator" which automates experiments to measure the phase behavior of proteins, thus enabling one to rationally design protein crystallization screens customized for the protein of interest.[62] We also developed a chip which enables one to grow protein crystals in a microfluidic environment and then use the chip in a synchrotron beam line to take diffraction data in situ.[63] Both of these approaches were able to substantially increase success rates in crystallization and quality of diffraction data.[64]

We became interested in using microfluidics to measure molecular interactions, which led us to develop a technology called Mechanically Induced Trapping of Molecular Interactions (MITOMI).[65] The principle of this approach is to use a deflectable membrane – very similar to the valve – which can trap molecular interactions on a glass surface. For example, if a transcription factor is anchored to the surface under the membrane, when the membrane is open and fluorescently labeled DNA is in solution, an equilibrium of bound and unbound DNA is reached. When the membrane is lowered, it traps the bound DNA against the surface and the unbound DNA can be washed away. If the remaining DNA happens to dissociate from the transcription factor, it will be retained against the surface by the membrane. In this fashion the off rate kinetics of the interaction are decoupled from the measurement, and it is possible to quantitate the bound DNA at one's leisure. We initially built a chip with 400 such measurement units, and later built a chip with 4,000 unit cells.[66] A first application of the MITOMI chip was to measure the free energy of binding of a transcription factor against all possible DNA sequences, from which we were able to show that it was sufficient to use a table of such energies along with informatics knowledge of the genome to correctly predict which genes were regulated by the transcription factor. We also were able to show that linear approaches to represent sequence motifs do not capture the full complexity of the sequence recognition properties of a transcription factor.[67] In later work, we went on to use the MITOMI chip for drug screening[68] and to measure protein-protein interactions.[69] The latter ability enabled us to perform a proteome-wide screen in staph aureus in order to discover putative function of unannotated genes.[70]

Microfluidics have also become a powerful tool to automate genetic analysis. As one example, digital PCR is a technique which in principle allows absolute quantitation of nucleic acids.[71] This method uses limiting dilution to make aliquots which contain on average less than one DNA molecule. By performing PCR on all of the aliquots and measuring which aliquots have product, one can calculate how many starting molecules were in the initial solution in such a way that the quantitation is completely decoupled from amplification efficiency. This solves an outstanding problem in nucleic acid measurement and corrects one of the drawbacks of conventional PCR, but was impractical to perform with conventional fluid manipulations. It was evident the microfluidics could solve this problem, and Fluidigm used microfluidic large scale integration to develop the first commercial digital PCR system, for which applications were developed by many groups around the world. There are applications in basic science, such as single cell gene expression[72,73], discovering gene-organism relationships in the termite gut microbiome[74] and discovery of virus-host relationships.[75] Digital PCR can be used for measurements in clinical genetics[76], cancer diagnostics,[77,78] as well as for measuring GMO contamination in food.[79,80] Eventually droplet microfluidics was used to develop second generation digital PCR products, now sold by Biorad, which have increased the throughput and scaling of such measurements.

Another example of microfluidic automation in genetic analysis is the development of PCR arrays.[81] We used microfluidic large scale integration to develop a reactor matrix, in which there are channels which form columns and rows, and a reactor exists at every vertex. This approach enables one to perform $N^2$ experiments with only 2N pipetting steps, and it solves the "world to chip" problem by using each pipetted reagent for at least N reactions. This idea was commercially developed by Fluidigm and their Dynamic Array now has a 96x96 matrix, enabling more than 9,000 experiments to be performed with less than 200 pipetting steps. Typically DNA templates are loaded in one axis and PCR primers are loaded on the other axis, and the machine performs real time quantitative PCR, enabling measurement of either gene expression or genotyping. This device has found many applications, ranging from managing salmon fisheries in Alaska[82] to single cell gene expression analysis.[83,84]

Single cell genomics is an application of microfluidics in biology which is currently receiving intense attention. This area has moved beyond the single cell PCR measurements described above to whole genome and transcriptome sequencing from single cells. We used microfluidic large scale integration to sequence the first single cell genome from an uncultivated environmental microbe,[85] and have continued to use this approach to analyze a number of uncultivated microbial genomes.[86–90] As sequencing costs decreased, we developed devices which enabled single cell human genome analysis, which we used to perform human genome haplotyping[91] as well as an analysis of recombination statistics and de novo mutations in human sperm cells.[92] Fluidigm developed a commercial product called the C1 based on our work, and this device is now used in laboratories around the world to perform single cell transcriptome and genome sequencing. In my group, among other applications, we have used this device to study the developing mouse lung,[93] adult human neurons,[94] cellular reprogramming,[95] and the clonal structure of cancer.[96] The flexibility of microfluidic large scale integration has enabled this system to be used for other single cell measurements, such as epigenetic analysis by ATAC-seq. David Weitz and collaborators used droplet microfluidics to develop a device for single cell transcriptome analysis[97,98] which can do larger numbers of cells but at lower transcriptome coverage, which has now reached the commercial market through the company 10x Genomics and which is enjoying a number of clever applications.[99,100]

## Conclusion

Microfluidics as a field has matured enormously since its founding nearly four decades ago. Early dreams of creating the biological equivalent of the integrated circuit have been realized through the development of microfluidic large scale integration, enabling one to fabricate a "lab on a chip" with tens of thousands of individual micromechanical valves, and similarly to fabricate chips which make millions of individual encapsulated droplet experiments. None of this would have been possible without the development of a revolutionary new fabrication technology known as soft lithography, which introduced new processes and materials. Microfluidic technologies have moved beyond academic laboratories and there is now a thriving industry for devices based around biological and biochemical measurement, with many companies in the space and aggregate sales of hundreds of millions of dollars per year. These tools have truly solved the "tyranny of pipetting" problem and have powered discoveries in numerous areas of biology.